\documentclass[preprint,superscriptaddress]{revtex4-2} 
\usepackage{siunitx}
\usepackage{xcolor}
\usepackage{graphicx}
\usepackage[normalem]{ulem}
\usepackage{braket}
\usepackage{amsmath}

\begin{document}

\title{Spatiotemporal focusing through a multimode fiber via time-domain wavefront shaping}

\author{Matthias C. Velsink}
\affiliation{MESA+ Institute for Nanotechnology, University of Twente, PO Box 217, 7500 AE Enschede, The Netherlands}
\affiliation{Presently with Advanced Research Center for Nanolithography (ARCNL), Science Park 106, 1098 XG Amsterdam, The Netherlands}

\author{Lyubov V. Amitonova}
\affiliation{LaserLaB, Department of Physics and Astronomy, Vrije Universiteit Amsterdam, De Boelelaan 1081, 1081 HV Amsterdam, The Netherlands}
\affiliation{Advanced Research Center for Nanolithography (ARCNL), Science Park 106, 1098 XG Amsterdam, The Netherlands}
\author{Pepijn W. H. Pinkse}
\affiliation{MESA+ Institute for Nanotechnology, University of Twente, PO Box 217, 7500 AE Enschede, The Netherlands}

\email{p.w.h.pinkse@utwente.nl}

\begin{abstract}
We shape fs optical pulses and deliver them in a single spatial mode to the input of a multimode fiber. The pulse is shaped in time such that at the output of the multimode fiber an ultrashort pulse appears at a predefined focus. 
Our result shows how to raster scan an ultrashort pulse at the output of a stiff piece of square-core step-index multimode fiber and in this way the potential for making a nonlinear fluorescent image of the scene behind the fiber, while the connection to the multimode fiber can be established via a thin and flexible single-mode fiber. The experimental results match our numerical simulation well.
\end{abstract}
\maketitle

\section{Introduction}
All-optical imaging via multimode fibers (MMF) has the potential to become the method of choice for imaging in confined spaces, combining the smallest access diameter with the highest NA \cite{amitonova2016, turtaev2018}. The most important application is minimally invasive endoscopy, but other use cases such as product inspection in an industrial setting are notable as well \cite{flusberg2005}. Multimode fibers support a large number of optical modes and hence transmit patterns from their input to their output facet. However, complex multimode interference makes it challenging to reconstruct the original input; bending of the MMF scrambles the multimode interference \cite{ploschner2015}. Different methods have already been investigated to overcome this, such as spatial wavefront shaping \cite{cizmar2011, leonardo2011}, machine learning \cite{borhani2018}, and compressive sensing \cite{amitonova2020}. The driving force behind these advances is the experimental ability to control light fields in complex media \cite{vellekoop2007,kubby2019,rotter2017}. Nowadays, MMFs are showing more and more promise for minimally-invasive endoscopic imaging \cite{cizmar2012,mezil2020,pikalek2019}. However, so far most MMF imaging methods are based on linear scattering or absorption  \cite{papadopoulos2012,xiong2016,ngom2018,defienne2020}.

In free-space microscopy, a plethora of special imaging modalities have been devised exploiting nonlinear imaging with ultrashort pulses. Despite its complexity, nonlinear microscopy has multiple advantages. Nonlinear methods reduce out-of-focus background and phototoxicity, allow to initiate highly localized photochemistry in thick samples, and provide optical sectioning that results in higher sensitivity and 3D imaging ability \cite{helmchen2005}. Considerable efforts have been put into the development of nonlinear endo-microscopy methods \cite{fu2007, andresen2013}. Unfortunately, combining ultrashort pulses with MMF imaging is non-trivial, as the modal interference and modal dispersion in a MMF results in a complex spatiotemporal output field \cite{redding2013}. Despite the fact that for a GRIN fiber this is not so much of an issue \cite{pikalek2019}, step-index MMFs can provide multiple advantages such as better mode mixing and larger NA.

Long-range spatio-temporal intensity correlations for an optical pulse propagating through a MMF have been studied \cite{xiong2019}. The temporal control of the averaged light intensity after a MMF, at the expense of the spatial pattern, has been shown \cite{mounaix2019}. Recently, several nonlinear optical imaging techniques through a single MMF probe have been demonstrated including two-photon excitation microscopy \cite{morales-delgado2015a, turcotte2020}, 3D microfabrication based on two-photon polymerization \cite{morales-delgado2017}, and coherent anti-Stokes Raman scattering (CARS) microscopy \cite{tragardh2019}. All these methods of nonlinear imaging require spatial-domain wavefront shaping and consequently control over many spatial modes on the MMF input.

Here we propose a new approach for imaging through a single MMF probe. We ‘focus’ light at any point on the distal fiber facet by using a single input mode utilizing light scrambling in a MMF, pulse shaping in time, and a nonlinear optical detection. Our system allows control over the position of a nonlinearly focused beam \textit{in space} on the MMF output facet by shaping an input pulse in a single spatial mode \textit{in time}. In contrast to other methods of nonlinear focusing and imaging through strongly scattering media \cite{katz2011,katz2014}, the proposed approach does not rely on spatial wavefront shaping. Controlling only a temporal shape on the single-mode input allows us to avoid the spatial control over the MMF input. This way of light control at the MMF output can also help to avoid the perturbation sensitivity of MMF-based imaging probes. Moreover, our method does not require a reference beam and/or measurements of a temporal profile as method of spatio-temporal focusing of an ultrafast pulse through a scattering medium as shown in \cite{mccabe2011}. To summarize, to the best of our knowledge, this paper is the ﬁrst to experimentally demonstrate grid scanning an ultrashort pulse over the output facet of a stiff piece of the MMF by temporally re-shaping the single-mode input pulse using nonlinear optical feedback.

\section{Theoretical description}
\begin{figure}[htbp]
	\centering
	\includegraphics[width=5in]{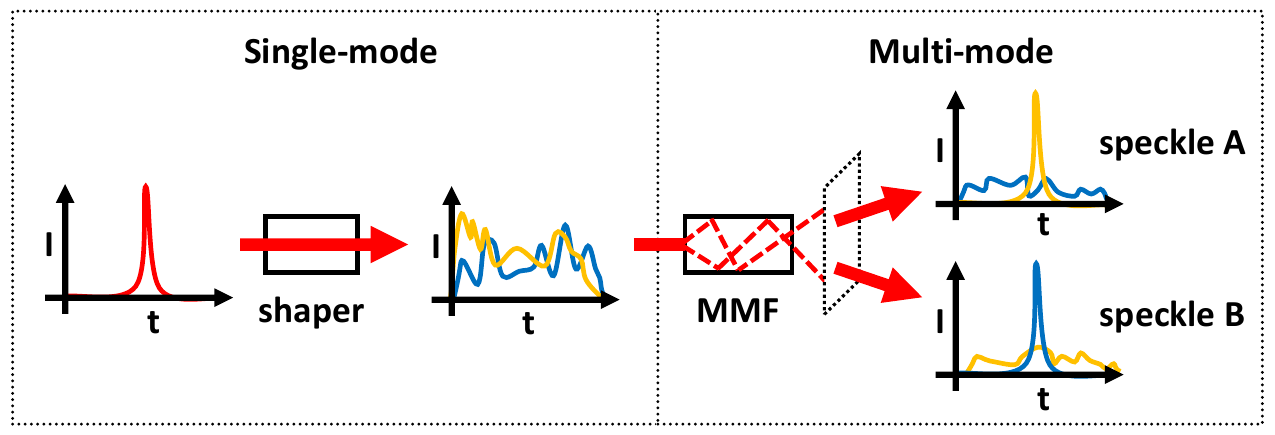}
	\caption{Sketch of the principle of time-domain wavefront shaping, which shows the intensity versus time traces of a pulse in different stages of the process. By shaping a transform-limited pulse (red trace) in time in a single spatial mode, either pulse A (yellow trace) or pulse B (blue trace) can be selected. These pulses will produce a transform-limited pulse in either spot A or spot B after travelling through a length of multimode fiber (MMF), but not in both spots simultaneously, because their response is independent. The corresponding pulse shapes are determined using time-domain wavefront shaping.}
	\label{fig:sketch}
\end{figure}

With continuous-wave (CW) light, it is possible to spatially shape the input field of a MMF in such a way that a focus appears at the output facet of the fiber.
However, time-domain shaping is necessary in order to allow the input to travel in a single spatial mode. The output field of a MMF for a broadband, pulsed input is also time-dependent, which can be exploited to do time-domain wavefront shaping. The principle is illustrated in Fig. \ref{fig:sketch}. Two spots at the output of the MMF, A and B, are assumed to have an independent temporal response to a transform-limited input pulse. Inverting one of the responses and using that as the input pulse shape results in a transform-limited pulse in either spot A or B, depending on which response was inverted. This enables making a short pulse at a particular spot at the output facet, even though the input pulse is still in a single spatial mode.
Note that all pulses travel via the same spatial path since they are injected via a single spatial mode, hence any losses affect all input pulses in the same way and effectively do not play a role.

We will now elaborate on why the output is time-dependent, how time-domain wavefront shaping is defined and how the system can be modelled.    

\subsection*{Time-dependent output patterns}
A multimode fiber supports many eigenmodes, each with its own spatial profile and propagation constant. For a focussed input field, many of these eigenmodes are excited simultaneously, with their amplitude given by the overlap integral of the mode field and the input field. The eigenmodes have different propagation constants. As a consequence, the light waves in the eigenmodes, characterized by their complex amplitudes, do not stay in phase after travelling through the fiber. As a result, the output field is a superposition of mode fields with seemingly random phases, leading to a speckled output field. The propagation constants and mode fields are in general also frequency-dependent, which leads to time-dependent fields inside the fiber and therefore also at the output facet. This time dependence we want to model here. Throughout this paper we ignore polarization. The time-dependent output field of a MMF of length $L$ with $N$ eigenmodes is described by
\begin{equation}
	E_{\text{out}}(x,y,t) = 
	\sum_{\omega}
	e^{-i\omega t}\sum_{n = 1}^{N}e^{i\beta_n(\omega)L}A_n(\omega)\Psi_n(x,y),
	\label{eq:general_output}
\end{equation}
where $\omega$ runs over discrete optical frequencies in the excitation light, $\beta_n(\omega)$ are the propagation constants, $A_n(\omega)$ the initial mode amplitudes, and $\Psi_n(x,y)$ the (orthonormal) fields of mode $n$. We consider a discrete set of frequencies for simplicity of the calculation, and we also ignore mode mixing. However, the full numerical simulation does include mode mixing, which is explained in Appendix B. Since the profiles of the eigenmodes are only weakly frequency-dependent, here we assume that they are completely frequency-independent for simplicity, although the simulation assumes a more general $\Psi_n(x,y,\omega)$. We take the input field to have a constant in-general-complex amplitude $C(x,y)$, but with a phase shift $\theta(\omega)$, so that $E_{\text{in}}(x,y,\omega) = \exp(i\theta(\omega))C(x,y)$. We can therefore approximate the  initial amplitudes of the eigenmodes with an overlap integral as
\begin{equation}
 	A_n(\omega) = e^{i\theta(\omega)}\int C(x^\prime,y^\prime)\Psi_n(x^\prime,y^\prime)dx^\prime dy^\prime \equiv e^{i\theta(\omega)}C_n.
	\label{eq:mode_amplitudes}
\end{equation}

\subsection*{Time-domain wavefront shaping}
By altering $\theta(\omega)$, we can change the output field in time and target a specific output location to produce an ultrashort pulse there. In the center of the fiber for example ($(x,y) = (0,0)$), the output field at $t = 0$ is given by Eqs. (\ref{eq:general_output}) and (\ref{eq:mode_amplitudes}):
\begin{equation}
	E_{\text{out}}(0,0,0) = 
	\sum_{\omega}
	e^{i\theta(\omega)}\sum_{n = 1}^{N}e^{i\beta_n(\omega)L}C_n\Psi_n(0,0).
\end{equation}
By setting the phase shifts to
\begin{equation}
	\theta(\omega) = -\arg\left(\sum_{n = 1}^{N}e^{i\beta_n(\omega)L}C_n\Psi_n(0,0)\right),
\end{equation}
we have
\begin{equation}
	E_{\text{out}}(0,0,0) = \sum_{\omega}\left|\sum_{n = 1}^{N}e^{i\beta_n(\omega)L}C_n\Psi_n(0,0)\right|,
\end{equation}
which is a strong peak due to all the contributions being in phase. In general, the argument (and the amplitude) of the inner sum in Eq. (\ref{eq:general_output}) varies rapidly with $x$ and $y$. Since $\theta(\omega)$ is fixed and independent of $x$ and $y$, the sum over all frequencies for positions away from $(x,y) = (0,0)$ is incoherent and the output is therefore not peaked in time there. To produce a peaked pulse in time at an arbitrary position $(x,y) = (X,Y)$, we can simply set
\begin{equation}
	\theta(\omega) = -\arg\left(\sum_{n = 1}^{N}e^{i\beta_n(\omega)L}C_n\Psi_n(X,Y)\right).
\end{equation}
In an experimental setting, however, the exact propagation constants are not known, and mode mixing further complicates the propagation through the fiber, so that the required phase shifts cannot be calculated a priori. Instead, the phase shifts can be optimized using an iterative algorithm.

\section{Experimental details}
\begin{figure}[htbp]
	\centering
	\includegraphics[width=4in]{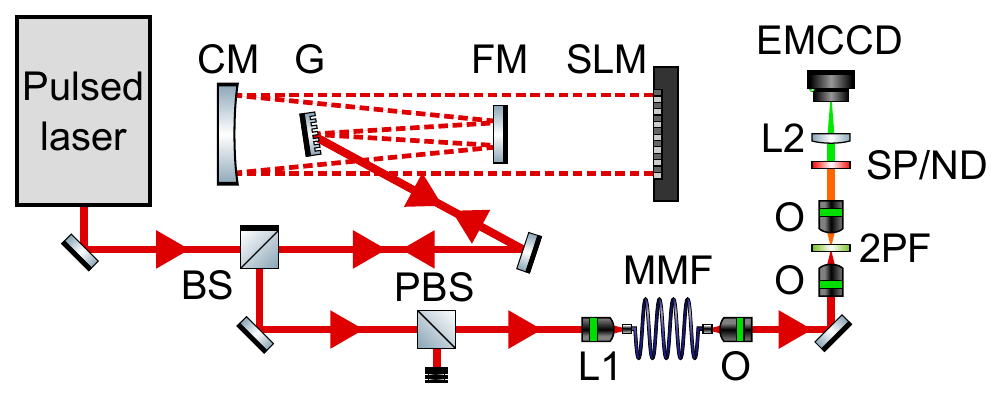}
	\caption{Schematic of the experimental setup. The output of the pulsed laser is directed to the pulse shaper, which consists of (polarizing) beam splitters ((P)BS), a grating (G), a cylindrical mirror and folding mirror (CM/FM), and a spatial light modulator (SLM). The shaped pulse is focussed into the multimode fiber (MMF) using an aspheric lens (L1, \SI{8}{\mm} focal length). The output light is collected, refocussed into a two-photon fluorescent screen (2PF), and again collected using matched objectives (O, 20X 0.4NA). A short-pass (SP) or neutral-density (ND) filter can be used for either nonlinear or linear imaging with a \SI{200}{\mm} lens (L2) and an electron-multiplying CCD camera (EMCCD).}
	\label{fig:setup_schematic}
\end{figure}

To experimentally verify the principle of time-domain wavefront shaping, we use the setup as illustrated in Fig. \ref{fig:setup_schematic}. The output of a mode-locked Ti:Sa laser with \SI{13}{\nm} bandwidth ($\approx$\SI{100}{\fs}) pulses, centred at \SI{800}{\nm} (Spectra Physics Tsunami, \SI{80}{\MHz}), is shaped in time with a $4f$ pulse shaper \cite{monmayrant2004}. The pulse shaper uses a 640-pixel linear spatial light modulator (CRI SLM-640-D-VN) with a spectral resolution of \SI{0.064}{\nm/pixel}. We have calibrated it using a spectrometer \cite{dopke2015}. The output of the pulse shaper is focussed into a multimode fiber, after which the average output power is \SI{50}{\mW}.

It is extremely challenging to measure the speckle-like output pattern directly on the relevant ultrashort (sub-ps) timescales. It would require an ultrafast streak camera or spatially scanning a SPIDER, FROG or SEA TADPOLE pulse characterisation technique \cite{bowlan2006}, which is not available in our laboratory. To still detect temporal behaviour indirectly, the output pattern is imaged with a nonlinear method. To this end, the output facet of the MMF is imaged into a \SI{50}{\um} thin cuvette filled with a two-photon fluorescent medium (Rhodamine 6G in ethylene glycol). A thinner \SI{20}{\um} cuvette would better match the Rayleigh length of the focus, but unfortunately, it became very hard to fill, close and mount in a stable way. The medium does not have linear absorption for \SI{800}{\nm} pump light, but can absorb two \SI{800}{\nm} photons and emit a green fluorescence photon \cite{makarov2008}. This two-photon process is sensitive to the square of the instantaneous optical power, so temporal compression can be made visible. In the future, it might be beneficial to select a single spatial output mode and detect that without the dye but using a nonlinear photodiode and lock-in detection instead, for an increased signal-to-noise ratio. The pump light is removed with a short-pass filter and the weak fluorescence is imaged with an EMCCD camera with high gain (Andor iXon DV885). Swapping the short-pass filter for an ND filter allows for linear imaging of the output intensity.

\subsection*{Square-core multimode fiber and numerical simulation}
 The ultrafast temporal behaviour of the shaped pulses and the output are of interest for the present study, but unfortunately we are unable to characterize this in our experiment. For this reason, we use a numerical simulation of the square-core fiber for testing time-domain wavefront shaping algorithms and to simulate the ultrafast temporal behavior. The details of the simulation can be found in Appendix B.

The multimode fiber in the experiment is a 70-by-70 micron square-core fiber (Ceramoptec, 0.22 NA). We have chosen a square-core fiber because it has a flatter intensity profile at the output and shows less correlation between the input location and the output pattern in comparison to round MMFs, which indicates more mode mixing and better excitation of higher-order modes \cite{sutherland2016}. However, the proposed approach will in principal work with any kind of MMF. 

\begin{figure}[htbp]
	\centering
	\includegraphics[width=5in]{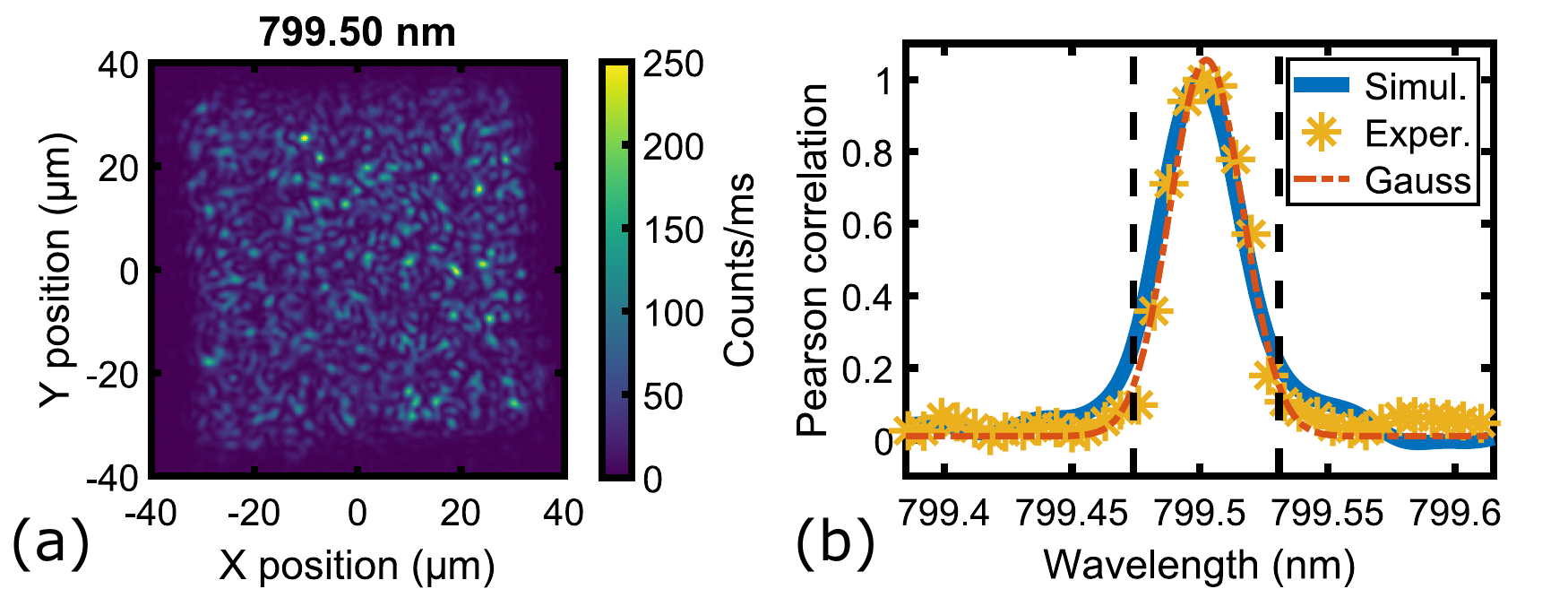}
	\caption{(a) Experimental linear image of the fiber output facet for a CW wavelength of \SI{799.50}{\nm}. The output intensity is clearly speckled due to modal dispersion and mode mixing. (b) Pearson correlation coefficient of the speckle pattern at \SI{799.50}{\nm} with speckle patterns measured at different wavelengths, for both the numerical simulation (the solid blue line) and the experiments (the yellow stars). Based on a Gaussian fit to the experimental data (the dashed red line), the speckle patterns decorrelate to a value of $1/e^2$ after a wavelength shift of about \SI{28}{\pm}, which is indicated on both sides with dashes black lines. The simulation closely matches the experimental data.}
	\label{fig:decorrelation}
\end{figure}

To characterize the frequency dependence of the output patterns, a tunable CW Ti:Sa laser (Coherent MBR-110) is used as input. The wavelength is scanned with a step size of \SI{0.1}{\per\cm} (\SI{7}{\pm}), limited by the resolution of our wavelength meter (Burleigh WA-10L). The experimental intensity pattern at \SI{799.50}{\nm} is shown in Fig. \ref{fig:decorrelation}(a). 
The Pearson correlation coefficients between the speckle pattern at 799.50\,nm and other speckle patterns as a function of wavelength are presented in Fig. \ref{fig:decorrelation}(b). Based on a Gaussian fit to the experimental data, the patterns decorrelate to a value of $1/e^2$ after wavelength shift of approximately \SI{28}{\pm}, which matches remarkably well with our zero-free-parameter simulation. 
The fact that the measured decorrelation width matches the simulation so well gives confidence that we understand the modal dynamics of this square-core fiber good enough to use it for our imaging method.

\subsection*{Optimization algorithm}
As explained in the theoretical section, in the experiment the phase shifts that correspond to an ultrashort pulse at a specific output position are difficult to determine a priori. Instead, the phase shifts are found with an optimization procedure. On the camera, a circle with an 8-pixel radius ($\approx$ \SI{2.5}{\um}  at the MMF output facet) is placed around the desired output position. Furthermore, a square of 180-by-180 pixels ($\approx$ \SI{58}{\um} at the MMF output facet) is placed for background intensity measurements. For clarity, these regions are indicated in Fig. \ref{fig:before_after}. In every step of the algorithm, 160 of the central 320 pulse shaper pixels are randomly selected and shifted relative to their current phase from $0$ to $2\pi$ in increments of $\pi/4$. At each shift, the average nonlinear intensities in the circle and in the square (excluding the circle) are recorded. Due to the low level of nonlinear signal, the optimization circle was chosen to be larger than a single diffraction-limited spot of our system. As a result, several diffraction-limited spots might be optimized at the same time, decreasing the contrast. To improve the contrast of the optimization, we optimize on the ratio between the nonlinear intensity inside and outside the circle. Only in the beginning we optimize the intensity directly and slowly change into contrast enhancement, by gradually changing the optimization parameter from intensity to ratio. This optimization parameter will vary sinusoidally with the phase shift of the selected 160 pixels. At each step, a sine is therefore fitted through the phase shift and optimization parameter curve in order to find the phase shift that maximizes the optimization parameter. Then, this phase shift is added to the phases of the selected 160 pixels and a new random set of 160 pixels is selected for the next optimization step. After 3000 steps, the optimization is almost fully based on the ratio. The optimization is terminated after 5000 steps. This algorithm is inspired by the random partitioning algorithm from spatial wavefront shaping, where it is known that this type of algorithm gives a good signal-to-noise ratio in determining the optimal phases, because multiple controls are modulated simultaneously \cite{vellekoop2008a}. The phases are initially set to random values, so that the algorithm is more likely to find a global optimum. Appendix C describes the optimization procedure in more detail and shows an example of the progress during an optimization run.

Currently, the method we use is sequential and a single focal spot can be optimized at a time. The current 9 hours optimization time for the 25 grid points in this proof-of-principle demonstration is long, but needs to be performed only once for a given fiber configuration. We think it would be possible to parallelize optimization by using a time-domain version of the spatial-domain transfer matrix characterization method  \cite{vellekoop2008a}.

\section{Results and discussion}
\begin{figure}[htbp]
	\centering
	\includegraphics[width=5in]{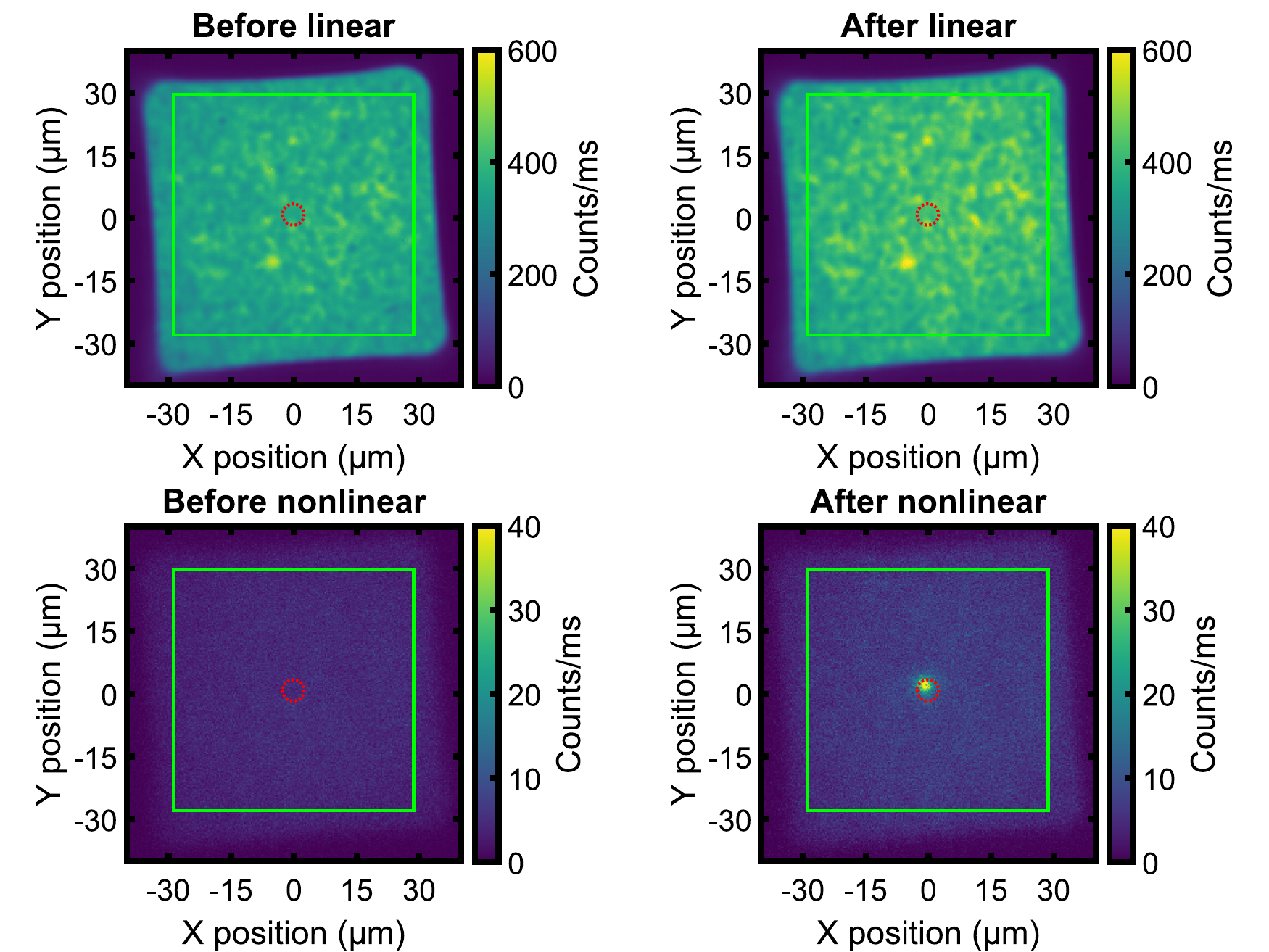}
	\caption{Experimental linear and nonlinear intensity images of the fiber output facet both before and after time-domain wavefront shaping in the center. The wavefront shaping region and the background region for the wavefront shaping algorithm are indicated by the dashed red circle and the green square, respectively. As expected, the linear intensity images show no difference, whereas the nonlinear intensity images show (nonlinear) focusing of light in a tight spot in the wavefront-shaping region.}
	\label{fig:before_after}
\end{figure}

Fig. \ref{fig:before_after} shows the results of an optimization run for a spot at the center of the fiber output facet. As expected for time-domain wavefront shaping, the linear imaging shows no difference between the before and after images. The nonlinear images, however, show clear focusing of light in the optimization region. This demonstrates that our optimization algorithm is able to find the optimal phase shift of the frequencies in the input pulse and can compress the output intensity in time very locally.

\subsection*{Enhancement and contrast}
An important parameter in wavefront shaping is the enhancement, defined as the ratio of (nonlinear) intensity in the wavefront shaping region after shaping and the (nonlinear)  intensity in the background \cite{vellekoop2008a}. To characterize the performance of our approach we now define the contrast as the enhancement minus one. Similar to (linear) spatial wavefront shaping, we expect that the contrast scales linearly with the number of controls in the pulse shaper and that the contrast should go to zero for zero controls. This expectation is perhaps counterintuitive, as the enhancement in our case can only be measured nonlinearly. But, even though the intensity peak in time grows quadratic with the number of frequencies and thus with the number of controls, the width of this peak shrinks linearly as well. This should result in a linear enhancement increase in a time-averaged nonlinear measurement after time-domain wavefront shaping. Mathematical support for the linear scaling of the enhancement can be found in Appendix A.

\begin{figure}[htbp]
	\centering
	\includegraphics[width=5in]{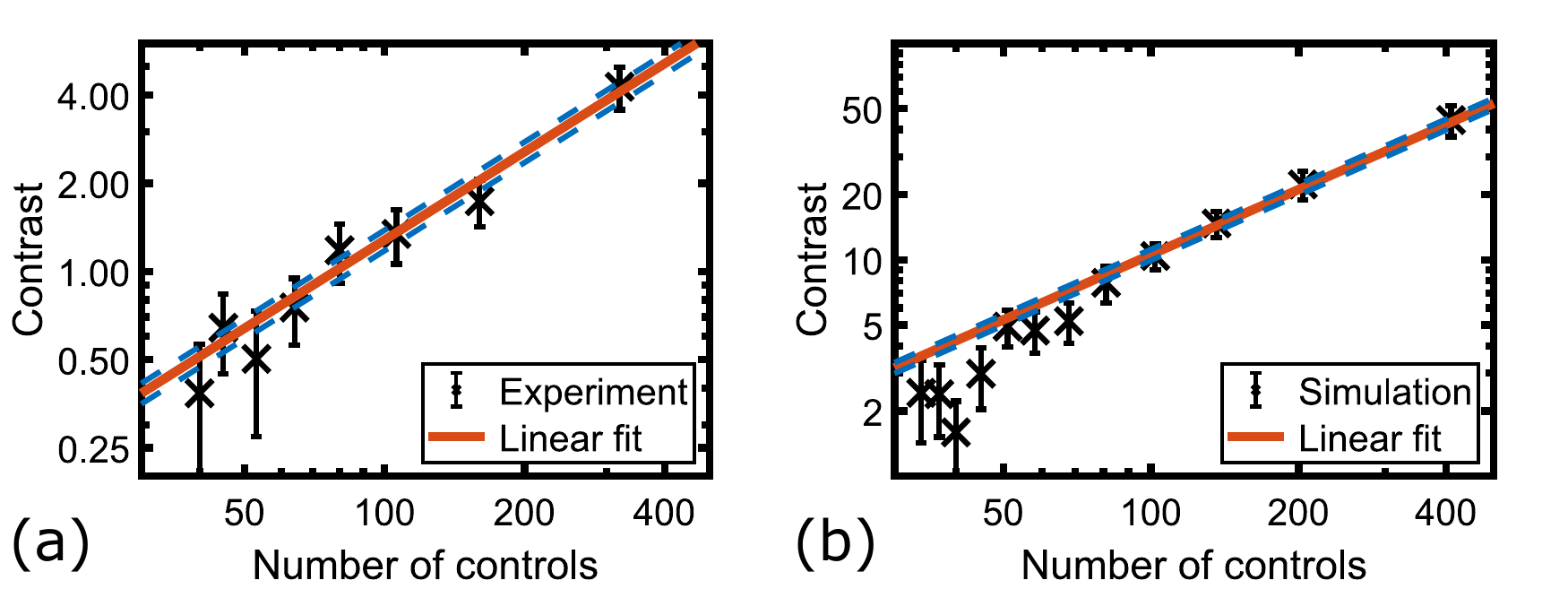}
	\caption{(a) Experimentally measured contrast as a function of the number of pulse-shaper controls. (b) Simulated contrast as a function of the number of pulse-shaper controls. Both the experiment and simulation indicate linear scaling with the number of controls. For a low number of controls, the simulated contrast falls below the linear fit. This could be due to an increased width of the enhanced spot, which causes nonlinear intensity to fall outside the averaging area used to determine the contrast. The error bars indicate the 95\% confidence interval in determining the contrast from a single image, and the blue dotted lines indicate the 95\% confidence interval in the linear fit.}
	\label{fig:enhancement}
\end{figure}

 By fitting a 2D Gaussian with offset to the bright spot in the wavefront shaping region after wavefront shaping, we find the center of the spot. The offset from the fit is used as the average intensity value of the background, $I_{\text{bg}}$. The average intensity of a 5-by-5 pixel region ($\approx$ \SI{1.5}{\um} at the MMF output facet), centered on the bright spot is also calculated as $I_{\text{bright}}$. The contrast $C$ is then calculated as $C = I_{\text{bright}}/I_{\text{bg}}$. We vary the effective number of controls by binning pulse-shaper pixels together and giving them the same phase, both in the experiment and in the simulation. The contrast plots are shown in Fig. \ref{fig:enhancement}.
By fitting a linear relationship without offset through the data (i.e., $C = aN_{\text{control}}$), it is clear that the contrast indeed grows linearly with the number of controls. The slopes from the fits are $0.013 \pm 0.001$ and $0.106 \pm 0.006$ for the experiment and simulation, respectively. For a low number of controls, the contrast from the numerical simulation falls below the linear fit. This might be explained by the observation that the optimized spot size in the simulation slightly grows with decreasing number of controls, which reduces the average intensity in the 5-by-5 pixel region and therefore also reduces the calculated contrast. The slope of the linear relationship is much lower in the experimental results, which we explain by the large amount of noise in the nonlinear imaging method. Noise is a combination of many different sources, such as shot noise, EMCCD noise, input power, pulse width, and temperature fluctuations and is difficult to quantify. Also, additional noise is added by out-of-focus nonlinear fluorescence in the cuvette. The effect of reduced contrast due to noise is also known in spatial wavefront shaping \cite{vellekoop2008a}. For the simulation, we simulate shot noise with a similar amplitude as in the experiment, but if we artificially increase the noise further it can happen that no enhancement is ever found, which further supports this reasoning. Despite the amplitude differences, the simulation confirms the linearity seen in both the experiment and the analytical theory. An advantage of the simulation is that we can achieve and explore regions of higher contrast, and confirm the linear scaling at contrast levels that are currently experimentally unreachable.

\subsection*{Temporal compression}
\begin{figure}[htbp]
	\centering
	\includegraphics[width=5in]{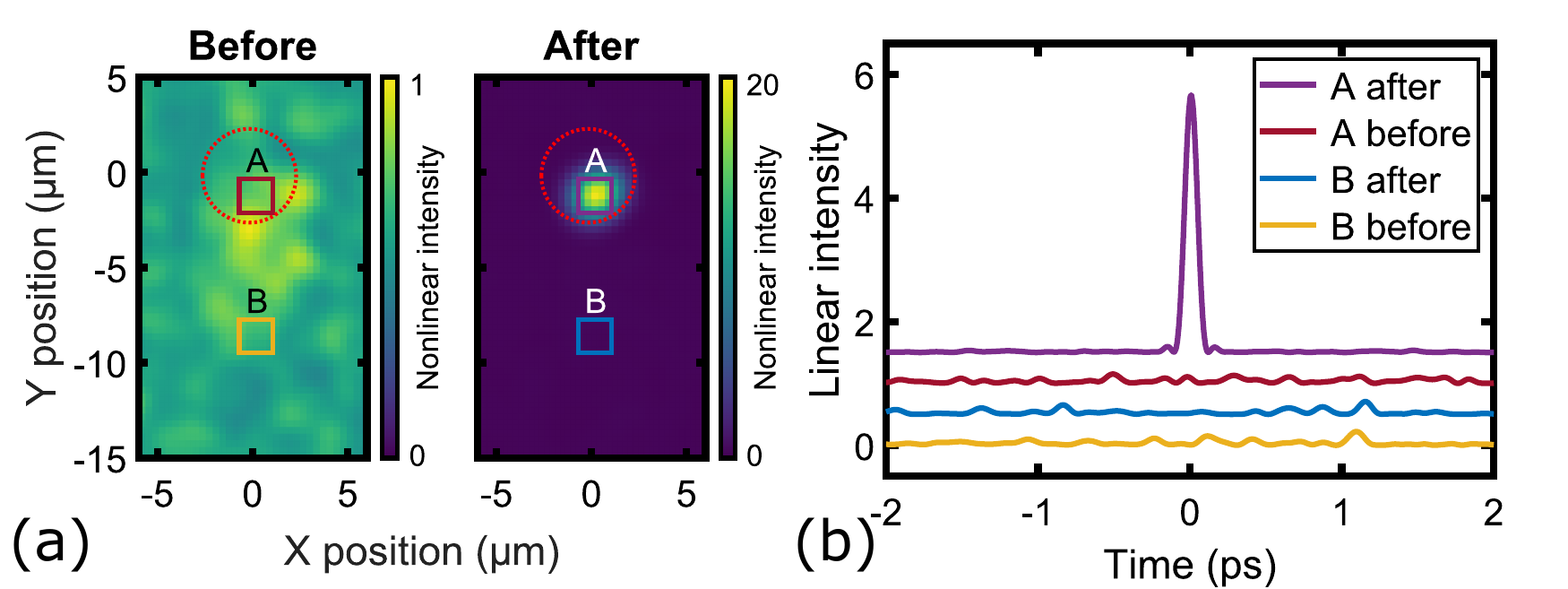}
	\caption{(a) Magnified simulated nonlinear image of the fiber output facet both before and after time-domain wavefront shaping in the center. The nonlinear intensity is normalized such that the maximum nonlinear intensity before optimization is 1. The wavefront shaping region is indication by the dashed red circle. Two 5-by-5 pixel regions, A and B, are highlighted. (b) Normalized linear intensity averaged over the regions A and B as a function of time before and after simulated time-domain wavefront shaping. Each trace is offset by 0.5 for clarity and all traces are scaled with a common factor such that the integral of the ``A after'' trace equals 1. Only the intensity in region A after shaping is sharply peaked in time.}
	\label{fig:sim_timetrace}
\end{figure}

As stated before, due to the multimode nature and very short timescale of the output intensity pattern, it is difficult to experimentally measure temporal compression at an output spot directly. However, we can use our numerical simulation and perform time-domain wavefront shaping with it. Fig. \ref{fig:sim_timetrace}(a) shows a magnification of the simulated nonlinear intensity both before and after time-domain wavefront shaping. For a time trace analysis, we choose two 5-by-5 pixel regions ($\approx$ \SI{1.5}{\um} at the MMF output facet), A and B, which are highlighted by the solid squares. The (normalized) average linear intensity in these square regions over a time period of \SI{4}{\ps}, both before and after optimization, is shown in Fig. \ref{fig:sim_timetrace}(b). Before optimization, both regions show a random and broad distribution of light in time, which is expected due to the random phase-shifts of the frequencies that are present at the spots. After optimizing in region A, however, region B still shows a similar trace but region A now shows a high and narrow pulse of light. This confirms the idea we sketched in Fig. \ref{fig:sketch}, namely that both spots have an independent temporal response, which can be selectively compressed by finding the corresponding optimal input pulse shape.

\subsection*{Raster scanning}
\begin{figure}[htbp]
	\centering
	\includegraphics[width=5in]{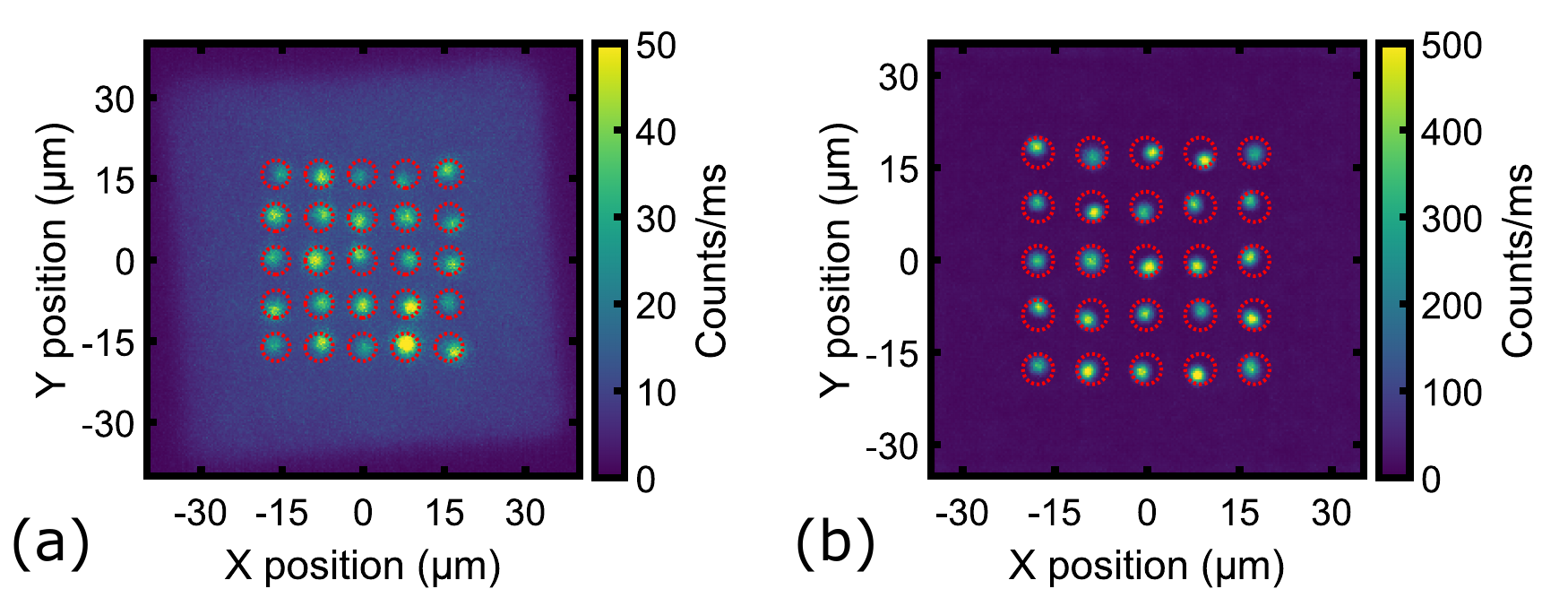}
	\caption{(a) Composite image of experimental time-domain wavefront shaping at 25 different positions (\SI{8}{\um} spacing). Each pixel value in this composite image is the maximum value over all seperate wavefront shaping runs. The dashed red circles indicate the wavefront shaping regions. This demonstrates the ability to raster scan an ultrashort pulse at the output facet of the MMF. (b) Again a composite image, but now using our numerical simulation. Intensity and position variations in the spots are similar to the experiment, but the typical contrast is an order of magnitude larger.}
	\label{fig:composite_grid}
\end{figure}

In order for time-domain wavefront shaping to have applications in nonlinear endoscopic imaging, a single optimization position is not sufficient. The simplest way to scan an ultrashort pulse over the entire output facet of the MMF is to define an optimization grid with many points and optimize the nonlinear intensity for each point individually. Fig. \ref{fig:composite_grid}(a) shows a composite image of such a grid after optimizing the input pulse shape for 25 points separately. Each pixel value in Fig. \ref{fig:composite_grid}(a) represents the maximum value over 25 recorded images obtained after sequential projections of the 25 optimized wavefronts. It is clear that not all points have the same intensity, and they also have slight positional variations with respect to the equidistant grid spacing we defined. We observe similar behaviour in our numerical simulation, for which a composite image with similar grid spacing is shown in Fig. \ref{fig:composite_grid}(b). Both the experimental and simulated figures are scaled by the experimental exposure time of \SI{1}{s}.

The variations in intensity contrast are likely due to noise in the nonlinear imaging method, which can make it difficult for the algorithm to precisely determine the optimal phase.  The experimental contrast varies between 3 and 8, with an average of 4.5, which is very comparable to the results presented in Fig. \ref{fig:enhancement}(a). The positional variations are likely due to the large optimization region we use in the algorithm. A spot can start to get enhanced anywhere in this region, which causes random variations in the final focus position. 

\section{Conclusion and outlook}
We have demonstrated spatial grid scanning of an ultrashort pulse at the output facet of a square-core multimode fiber by only changing the temporal shape of an ultrashort pulse at the input facet. The results match well with our numerical simulation, which can be used to directly show temporal behaviour at the ultrashort timescales. The current long optimization time is unpractical. However, if the MMF is stiff, the different shapes can be stored and reused many times, so that the  optimization time needs to be spent only once. The spectacular property of the proposed approach of nonlinear imaging through a MMF is the ability to control a \textit{spatial} position of the focus spot on the MMF output by using a single spatial mode at the input. As a result, a single-mode fiber, which is insensitive to spatial perturbations, can be used for endoscopic delivery of the input pulse through a flexible probe. Moreover, the fluorescence from a single grid position can in principle be collected back through the same MMF. One can combine a long flexible single-mode fiber with a rigid piece of a MMF to create a single semi-flexible fiber probe.

To summarize, the proposed approach of single-mode time-domain wavefront shaping enables deterministic and robust grid scanning of an ultrashort pulse over the fiber output facet. It paves the way toward the design of a flexible high-resolution nonlinear imaging probe and potentially has many applications in endoscopic bioimaging.

\section*{Appendix A: Enhancement derivation}\label{app:enhancement}
We first analytically calculate the enhancement in time-domain wavefront shaping for a general frequency-dependent transmissive medium.
Ordinary scattering media are typically modeled with $N$ spatial input modes and a complex-valued transmission matrix $t_{mn}$ that connects the field of the $n^{\text{th}}$ input mode to the $m^{\text{th}}$ output mode \cite{popoff2010,vellekoop2007}. With the $n^{\text{th}}$ input field written as $E_n^{\text{in}} = A_ne^{i\phi_n}$, we have
	\begin{equation}\label{eq:scattering}
		E_m^{\text{out}} = \sum_{n = 1}^{N}t_{mn}E_n^{\text{in}} = \sum_{n = 1}^{N}t_{mn}A_ne^{i\phi_n}. 
	\end{equation} 
Our model for a time-domain medium is analogous to this. We take a single mode as input, and $M$ spatial output modes. Furthermore, we assume a discrete set frequencies $\Omega$ of size $N$ and spacing $\delta$.
	\begin{equation}\label{eq:frequencyspace}
		\Omega = \{\omega_0,\omega_0 + \delta,...,\omega_0 + (N - 1)\delta\}.
	\end{equation}
Again, we use a complex matrix $t_{mn}$ to connect the input field of the $n^{\text{th}}$ frequency mode to the $m^{\text{th}}$ spatial output mode. With the input field of the $n^{\text{th}}$ frequency mode as $E_n^{\text{in}}(t) = A_ne^{i[(\omega_0 + n\delta)t + \phi_n]}$, the output field in the $m^{\text{th}}$ spatial mode is given by
	\begin{equation}\label{eq:frequencyscattering}
		E_m^{\text{out}}(t) = \sum_{n = 0}^{N-1}t_{mn}E_n^{\text{in}} = \sum_{n = 0}^{N-1}t_{mn}A_ne^{i[(\omega_0 + n\delta)t + \phi_n]}. 
	\end{equation} 
Both the input field and output fields are time dependent and $2\pi/\delta$-periodic. For each $\omega \in \Omega$, we assume the transmission to be independent and random in phase for each output mode. For simplicity, we consider only random phase, but fixed amplitude for each spatial output mode. Furthermore, the total transmission is taken to be unity. Under these assumptions,
	\begin{equation}\label{eq:frequencytransmission}
		t_{mn} = \frac{1}{\sqrt{M}}e^{i\theta_{mn}}\text{, with } 
		\theta_{mn} \in \left[0,2\pi\right)\text{ and } 
		f_{\Theta_{mn}}(\theta_{mn}) = \frac{1}{2\pi}, 
	\end{equation} 
where $f_{\Theta_{mn}}(\theta_{mn})$ is the probability density function for $\theta_{mn}$. 

It is easy to see from Eq. (\ref{eq:scattering}) that when the input is phase shaped such that $\phi_n = -\arg(t_{mn})$, the amplitude of $E_m^{\text{out}}$ is maximized. This is called wavefront shaping \cite{vellekoop2007}. Similarly, we can maximize $E_m^{\text{out}}(t=0)$ by setting $\phi_n = -\arg(t_{mn}) = -\theta_{mn}$, as then all the frequency components are in phase (see Eq. (\ref{eq:frequencyscattering})). Since the input field is in a single spatial mode and only depends on time, we call this time-domain wavefront shaping.

In wavefront shaping, the most important figure of merit is the enhancement $\eta$, which is defined as the ratio of the intensity in the shaping region after optimization, $I_N$, and the intensity in the same region with the same optimized input, but ensemble-averaged over all possible samples, $\left<I_0\right>$. So,	
	\begin{equation}\label{eq:enhancement}
		\eta = \frac{I_N}{\left<I_0\right>},
	\end{equation}
where $\left<...\right>$ denotes the ensemble-averaged expected value. In spatial wavefront shaping, assuming circular complex Gaussian random $t_{mn}$ \cite{vellekoop2007}, 
	\begin{equation}\label{eq:enhancementspatial}
		\eta = \frac{\pi}{4}(N - 1) + 1.
	\end{equation}
The enhancement thus scales linearly with the number of controlled modes $N$.

We now derive the enhancement for time-domain wavefront shaping using the model defined above. Using Eq. (\ref{eq:frequencytransmission}), we find three useful ensemble-averaged expected values:
	\begin{equation}\label{eq:averages}
		\begin{aligned}
		\left<t_{mn}\right> &= \frac{1}{\sqrt{M}}\frac{1}{2\pi}\int_{0}^{2\pi}e^{i\theta_{mn}}d\theta_{mn} = 0, \\ 
		\left<|t_{mn}|\right> &= \frac{1}{\sqrt{M}}\frac{1}{2\pi}\int_{0}^{2\pi}d\theta_{mn} = \frac{1}{\sqrt{M}}, \\
		\left<t_{mn}t_{m'n'}^*\right> &= \frac{1}{M}\frac{1}{(2\pi)^2}\int_{0}^{2\pi}\int_{0}^{2\pi}e^{i(\theta_{mn} - \theta_{m'n'})}d\theta_{mn}d\theta_{m'n'} = \frac{1}{M}\delta_{mm'}\delta_{nn'}.
		\end{aligned}
	\end{equation}
Without any phase conjugation and with unity amplitude input $E_n^{\text{in}}(t) = e^{i(\omega_0 + n\delta)t}$, the expected value for the output intensity in the $m^{\text{th}}$ output mode is given by
	\begin{equation}\label{eq:backgroundtime}
		\begin{aligned}
		\left<I_m^{\text{out}}(t)\right> &= \left<|E_m^{\text{out}}(t)|^2\right> = \left<\left|\sum_{n= 0}^{N - 1}e^{in\delta t}t_{mn}\right|^2\right> = \sum_{n,n'= 0}^{N - 1}e^{i(n-n')\delta t}\left<t_{mn}t_{mn'}^*\right> \\
		&= \sum_{n,n'= 0}^{N - 1}e^{i(n-n')\delta t}\frac{1}{M}\delta_{nn'} = \frac{N}{M}.
		\end{aligned}
	\end{equation}
If we phase conjugate for output mode $j$, then $E_n^{\text{in}}(t) = e^{i[(\omega_0 + n\delta)t - arg(t_{jn})]} = \sqrt{M}t_{jn}^*e^{i(\omega_0 + n\delta)t}$. For any output mode $k \neq j$,
	\begin{equation}\label{eq:othermodetime}
		\begin{aligned}
		\left<I_k^{\text{out}}(t)\right> &= \left<|E_k^{\text{out}}(t)|^2\right> = M\left<\left|\sum_{n= 0}^{N - 1}e^{in\delta t}t_{kn}t_{jn}^*\right|^2\right> \\
		&= M\sum_{n,n'= 0}^{N - 1}e^{i(n-n')\delta t}\left<t_{kn}t_{kn'}^*t_{jn}^*t_{jn'}\right> \\
		&= M\sum_{n,n'= 0}^{N - 1}e^{i(n-n')\delta t}\frac{1}{M^2}\delta_{nn'} = \frac{N}{M},
		\end{aligned}
	\end{equation}
where we have used the fact that $t_{kn}$ and $t_{jn}$ are independent for $k \neq j$. It is logical that Eqs. (\ref{eq:backgroundtime}) and (\ref{eq:othermodetime}) are the same if all $t_{mn}$ are uncorrelated, since averaging with an unshaped input wavefront is then the same as averaging with a random input wavefront. For the output mode $j$, we find
	\begin{equation}\label{eq:focusmodetime}
		\begin{aligned}
		\left<I_j^{\text{out}}(t)\right> &= \left<|E_j^{\text{out}}(t)|^2\right> = M\left<\left|\sum_{n= 0}^{N - 1}e^{in\delta t}t_{jn}t_{jn}^*\right|^2\right> = \frac{1}{M}\left|\sum_{n = 0}^{N - 1}e^{in\delta t}\right|^2 &= \frac{1}{M}\left|\frac{1 - e^{iN\delta t}}{1 - e^{i\delta t}}\right|^2 \\
		&= \frac{1}{M}\frac{\sin^2(N\delta t/2)}{\sin^2(\delta t/2)}.
		\end{aligned}
	\end{equation}
Based on the definition of the enhancement and using Eqs. (\ref{eq:othermodetime}) and (\ref{eq:focusmodetime}), we write
	\begin{equation}\label{eq:intensitiestime}
	\begin{aligned}
		\left<I_0(t)\right> &= \left<I_k^{\text{out}}(t)\right> = \frac{N}{M}\\
		\left<I_N(t)\right>	&= \left<I_j^{\text{out}}(t)\right> = \frac{1}{M}\frac{\sin^2(N\delta t/2)}{\sin^2(\delta t/2)}.
	\end{aligned}
	\end{equation} 
Since $\left<I_N(t)\right> = N^2/M$ for $t \to 0$, the maximum (ensemble-averaged) enhancement in time is $N$. However, the temporal features in the input and output fields are of the order $\Delta t \sim 1/N\delta$. For a physical system, the total bandwidth $N\delta$ can be several THz, giving temporal features in the femtosecond regime. As said before, this makes a direct, time-resolved measurement of $\left<I_N(t)\right>$ very hard. A physical detector will thus likely measure a time-averaged signal. To emulate a time-averaged measurement, we can time-average over a single period $2\pi/\delta$, because the input and output fields are periodic in time. A linear detector will detect signals proportional to
	\begin{equation}\label{eq:lineartime}
		\begin{aligned}
		S_0^1 &= \int_{-\pi/\delta}^{\pi/\delta}\left<I_0(t)\right>dt = \int_{-\pi/\delta}^{\pi/\delta}\frac{N}{M}dt = \frac{2\pi}{\delta}\frac{N}{M} \\
		S_N^1 &= \int_{-\pi/\delta}^{\pi/\delta}\left<I_N(t)\right>dt = \int_{-\pi/\delta}^{\pi/\delta}\frac{1}{M}\frac{\sin^2(N\delta t/2)}{\sin^2(\delta t/2)}dt = \frac{2\pi}{\delta}\frac{N}{M}.
		\end{aligned}
	\end{equation}
Both signals are the same, hence it is impossible to perform time-domain wavefront shaping with the feedback of a time-averaged linear detector. Because we can only shape an input pulse in the time domain, we cannot increase the average output energy in a spatial output mode. It is possible to use a linear detector and an interferometric measurement to reconstruct the output fields, but this method is slow and therefore not suitable for direct feedback for wavefront shaping.

A possibility to get a feedback signal to base our temporal wavefront shaping on is the use of a {\em non-linear} detector. Let's assume such a detector is sensitive to $I^2(t)$, allowing to detect signals proportional to
	\begin{equation}\label{eq:nonlineartime}
		\begin{aligned}
		S_0^2 &= \int_{-\pi/\delta}^{\pi/\delta}\left<I_0(t)^2\right>dt = \int_{-\pi/\delta}^{\pi/\delta}\frac{N^2}{M^2}dt = \frac{2\pi}{\delta}\frac{N^2}{M^2} \\
		S_N^2 &= \int_{-\pi/\delta}^{\pi/\delta}\left<I_N(t)^2\right>dt = \int_{-\pi/\delta}^{\pi/\delta}\frac{1}{M^2}\frac{\sin^4(N\delta t/2)}{\sin^4(\delta t/2)}dt = \frac{2\pi}{\delta}\frac{2N^3 + N}{3M^2}.
		\end{aligned}
	\end{equation}
Here, we have used that for any output mode $k \neq j$,
    \begin{equation}\label{eq:othermodetime2}
		\begin{aligned}
		\left<I_k^{\text{out}}(t)^2\right> &= \left<|E_k^{\text{out}}(t)|^4\right> = M^2\left<\left|\sum_{n= 0}^{N - 1}e^{in\delta t}t_{kn}t_{jn}^*\right|^4\right> \\
		&= M^2\sum_{n,n',m,m'= 0}^{N - 1}e^{i(n-n'+m-m')\delta t}\left<t_{kn}t_{kn'}^*t_{km}t_{km'}^*t_{jn}^*t_{jn'}t_{jm}^*t_{jm'}\right> \\
		&= M^2\sum_{n,n',m,m'= 0}^{N - 1}e^{i(n-n'+m-m')\delta t}\frac{1}{M^4}\delta_{nn'}\delta_{mm'} \\
		&= \frac{1}{M^2}\sum_{n,m= 0}^{N - 1}1 = \frac{N^2}{M^2},
		\end{aligned}
	\end{equation}
and that for the output mode $j$,
    \begin{equation}\label{eq:focusmodetime2}
		\begin{aligned}
		\left<I_j^{\text{out}}(t)^2\right> &= \left<|E_j^{\text{out}}(t)|^4\right> = M^2\left<\left|\sum_{n= 0}^{N - 1}e^{in\delta t}t_{jn}t_{jn}^*\right|^4\right> = \frac{1}{M^2}\left|\sum_{n = 0}^{N - 1}e^{in\delta t}\right|^4 &= \frac{1}{M^2}\left|\frac{1 - e^{iN\delta t}}{1 - e^{i\delta t}}\right|^4 \\
		&= \frac{1}{M^2}\frac{\sin^4(N\delta t/2)}{\sin^4(\delta t/2)}.
		\end{aligned}
	\end{equation}
With this detector, the (ensemble-averaged) enhancement thus becomes
	\begin{equation}\label{eq:enhancementtime}
		\eta = \frac{S_N^2}{S_0^2} = \frac{2}{3}\left(N + \frac{1}{2N}\right).
	\end{equation}
The enhancement expression is very similar to the result for spatial wavefront shaping (equation (\ref{eq:enhancementspatial})). For large $N$, $\eta \approx 2N/3$, so the enhancement scales  linearly with $N$. Since $\left<I_N(0)\right>^2/N^2 \propto N^2$, one might expect the enhancement to grow quadratic and not linear. However, the width of the central peak must decrease linearly with $N$ because of energy conservation, so $S_N^2/S_0^2$ is linear in $N$. This effect is well-known in non-linear detection, where for constant average power the signal scales inversely proportional to the pulse width \cite{tang2006}. Our experimental implementation of a non-linear detector is discussed in the main paper.

So far we have assumed the time-domain medium to be loss free. In case it is lossy, the modeling assumption that all $|t_{mn}| = 1/\sqrt{M}$ will no longer be valid. Making the amplitude of $t_{mn}$ also randomly distributed will change the average values from Eq. (\ref{eq:averages}). This may alter the theoretical enhancement for small $N$, but for large $N$ the enhancement should still be linear in $N$, which is the most important result of this section. 

A complication in the experiment is that we cannot naturally vary the true number of independent frequency channels, as the resolution of our pulse shaper is fixed. Instead, we emulate smaller $N$ by binning together pixels on the pulse shaper SLM. This does not result in a lower number of frequency channels in the sample, but does result in fewer controllable frequency channels, effectively reducing $N$.

\section*{Appendix B: Numerical simulation}\label{app_numerical}
In the following we detail our numerical simulation of the square-core fiber. We will first describe the transverse fiber modes, then the propagation through the fiber and finally discuss the parameter choices to mimic the actual fiber used in the experiment.
\subsection*{Mode profiles}
We only consider a single polarization (horizontal) in the square-core fiber, for which one component of the electric field is given by 
\begin{equation}\label{eq:squarefield}
	E_x (p,q) = A\cos{\left[k_x(p)\,x - \phi(p)\right]}\cos{\left[k_y(q)\,y - \psi(q)\right]},
\end{equation}
  where we follow Ref. \cite{okamoto2006} throughout. Here,
\begin{equation}\label{eq:squarephases}
	\begin{aligned}
	\phi &= (p - 1)\frac{\pi}{2},\\
	\psi &= (q - 1)\frac{\pi}{2},
	\end{aligned}
\end{equation}
for mode numbers $p$ and $q$ (both $1,2,...$). Note that the coordinate system origin is centered on the square-core fiber. The wave numbers can be found with the transcendental equations
\begin{equation}\label{eq:squarewavenumbers}
	\begin{aligned}
	k_x \,a &= (p - 1)\frac{\pi}{2} + \tan^{-1}\left(\frac{n_{\rm co}^2\,\gamma_x}{n_{\rm cl}^2\,k_x}\right),\\
	k_y \,a &= (q - 1)\frac{\pi}{2} + \tan^{-1}\left(\frac{\gamma_y}{k_y}\right),
	\end{aligned}
\end{equation}
and
\begin{equation}\label{eq:squarewavenumbers2}
	\begin{aligned}
	\gamma_x^2 &=  k^2(n_{\rm co}^2 - n_{\rm cl}^2) - k_x^2,\\
	\gamma_y^2 &=  k^2(n_{\rm co}^2 - n_{\rm cl}^2) - k_y^2.
	\end{aligned}
\end{equation}
Here, $a$ is half the width of the fiber (i.e. the fiber is $2a$-by-$2a$), and $n_{\rm co}$ and $n_{\rm cl}$ are the core and cladding refractive index, respectively, and $k$ the wave number of the light. Eqs. (\ref{eq:squarewavenumbers}) and (\ref{eq:squarewavenumbers2}) are easily solvable using a few iterations of Newton's method, and the maximum values for $p$ and $q$ are those that still give a real solution for all wave numbers. Having found all possible values for $p$ and $q$, there are a total of $\max(p)*\max(q)$ modes, which can be enumerated with a single index $n$.  Combining Eqs. (\ref{eq:squarefield}) to (\ref{eq:squarewavenumbers2}) then gives the mode profiles $\Psi_n(x,y,\omega)$ as used in the main text. Finally, the propagation constant $\beta$ inside the fiber can be found with 
\begin{equation}\label{eq:squarepropagation}
    \beta^2 = k^2n_{\rm co}^2 - (k_x^2 + k_y^2).
\end{equation}
It is precisely this wavelength-dependent propagation constant that gives rise to phase shifts and independent speckle patterns for a broadband input.
\subsection*{Mode propagation and mode mixing}
Without any coupling between the modes, the evolution of the mode amplitudes $c_n$ is described by the matrix equation
\begin{equation}\label{eq:squarematrix}
    \frac{dc_n}{dz} = i\beta_nc_n.
\end{equation}
The solution of this equation for all modes and frequencies essentially results in Eq. (\ref{eq:general_output}). The initial amplitudes are found by calculating the overlap integral of the mode electric field profiles and the input electric field. 

In a real multimode fiber there is mode coupling between the modes due to bending and refractive index variations \cite{ploschner2015}. The coupling due to a single bend with radius $r$ may be described by
\begin{equation}\label{eq:squarematrixbend}
    \frac{dc_n}{dz} = i\beta_nc_n - \frac{n_{\rm co}k\xi}{r}\sum_mA_{nm}c_m,
\end{equation}
where $\xi$ is a correction factor (0.77 for silica), and 
\begin{equation}\label{eq:squareoverlap}
    A_{nm} = \braket{E_{x}(n)|\cos(\theta)x + \sin(\theta)y|E_{x}(m)}
\end{equation}
are the overlap integrals between modes $n$ and $m$ for a bend with projected angle $\theta$ with respect to the $x$-axis.
Eq. (\ref{eq:squarematrixbend}) can be solved as a matrix differential equation by using an eigenvalue decomposition.

\subsection*{Implementation}
Our fiber is simulated as a 1 meter long, 70-by-70 micron square-core fiber, with a numerical aperture of 0.22 and a cladding refractive index of 1.4533 (assuming pure silica at 800 nm light). No dispersion of the numerical aperture or refractive index is taken into account, only dispersion due to Eq. ($\ref{eq:squarepropagation}$). The pulse is modelled with a \SI{13}{nm} bandwidth ($\approx \SI{75}{fs}$ if transform-limited), with 2048 discrete frequencies from \SI{787}{nm} to \SI{813}{nm}. The modal dispersion alone stretches the pulse to around \SI{20}{ps}. The spatial coordinates are discretized with a resolution that is similar to the resolution of the experimental nonlinear imaging method, $\sim\SI{0.32}{\um}$ per pixel.

\begin{figure}[htbp]
	\centering
	\includegraphics[width=5in]{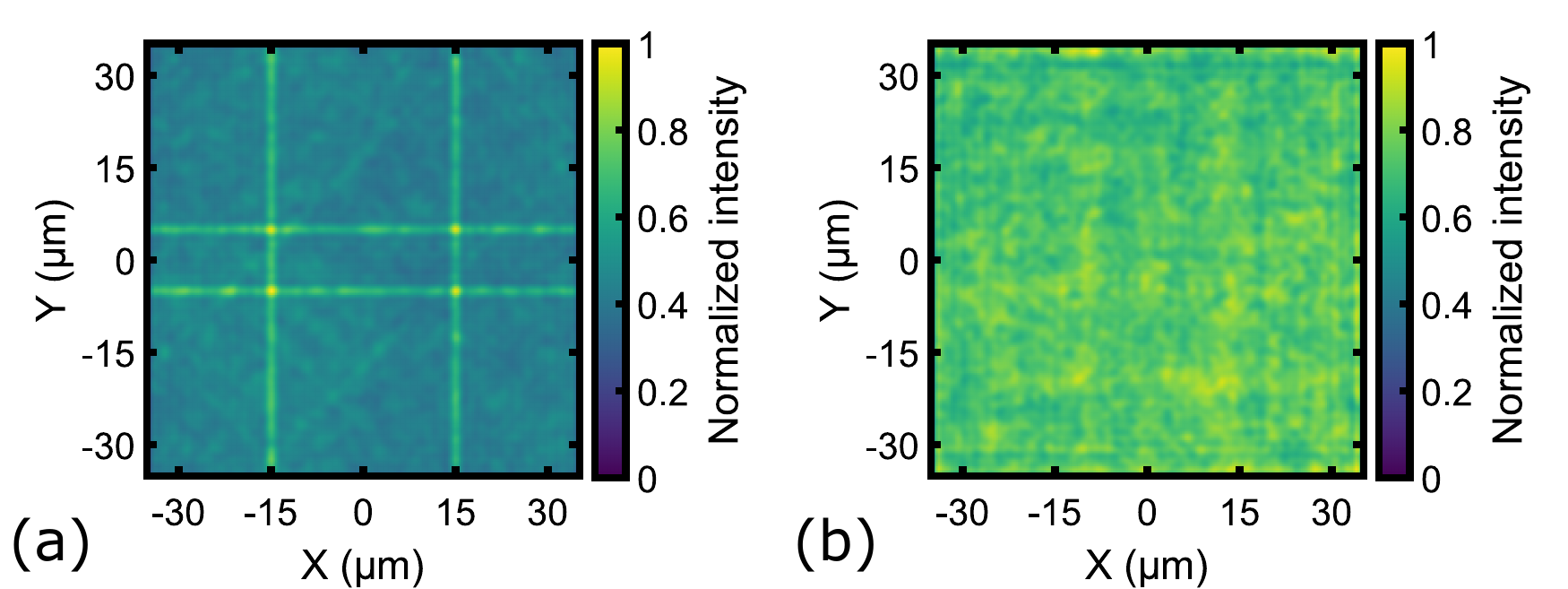}
	\caption{(a) Linear output intensity for a 0.22 NA diffraction limited input pulse at \SI{800}{\nm} with \SI{13}{\nm} bandwidth for the fiber simulation without coupling due to bends. The input location is at $x = \SI{15}{\micro\meter}$, $y = \SI{5}{\micro\meter}$. (b) Linear output intensity for the same input pulse, but now for the fiber simulation with coupling. The speckle pattern intensity is much more even due to mode coupling.}
	\label{fig:linear}
\end{figure}

In the lab, the fiber is strongly bent by winding it around a cylinder a few times and tie-wrapping it for stability. Because the exact bending is unknown, we randomly bend the simulated fiber 10 times, in a random direction and with a random radius of curvature between 1 and 4 centimeter. Note that the induced extra mode mixing was used to prove that our method can cope with that. It is not actually needed for the method to work, although it might help.
Table \ref{tab:bendparameters} shows the bend parameters in the final simulation. All bends are 10 centimeters long. For each bend, the propagation from beginning to end is calculated by solving Eq. (\ref{eq:squarematrixbend}). Fig. \ref{fig:linear} shows the difference for the linear output intensity for an ultrashort input pulse for the fiber simulation with and without mode coupling due to bending. Without bends, the output intensity shows intensity peaks at the input pulse location, because the modes are not coupled while propagating through the fiber. In contrast, the output intensity with bends is much more evenly distributed, without clearly visible patterns. A similar improvement is visible in the nonlinear output intensity.

\begin{table}[htbp]
\centering
\caption{\bf Bend parameters of the simulation}
\begin{tabular}{|ccc|ccc|}
\hline
Bend & Radius $r$ & Angle $\theta$ & Bend & Radius $r$ & Angle $\theta$ \\
\hline
 1 & \SI{1.82}{cm} & \ang{86.4}  &  6 & \SI{3.58}{cm} & \ang{281.5}\\
 2 & \SI{1.86}{cm} & \ang{257.7} &  7 & \SI{3.50}{cm} & \ang{330.3}\\
 3 & \SI{1.24}{cm} & \ang{177.1} &  8 & \SI{1.89}{cm} & \ang{339.9}\\
 4 & \SI{2.08}{cm} & \ang{4.2}   &  9 & \SI{3.45}{cm} & \ang{15.9} \\
 5 & \SI{3.04}{cm} & \ang{261.6} & 10 & \SI{3.02}{cm} & \ang{198.0}\\
\hline
\end{tabular}
  \label{tab:bendparameters}
\end{table}

\subsection*{Pulse shaper simulation}
To simulate the experimental pulse shaper, we divide the array of (complex) input frequency amplitudes into bins with the same spectral width as in the experiment (\SI{63}{\pm}). The phases of these bins can be changed during time-domain wavefront shaping. The frequency amplitude array with square bins is then convoluted with a Gaussian with a FWHM of $\sqrt{0.61}$ bins in order to simulate the finite width of a single frequency in the Fourier plane of our experimental pulse shaper (\SI{61}{\micro\meter} compared to \SI{100}{\micro\meter} wide pixels). The algorithm for wavefront shaping in the simulation is the same as the algorithm that is used in the experiment.

\section*{Appendix C: Time-domain wavefront shaping algorithm}\label{app:algorithm}
As explained in the main text, the wavefront shaping algorithm uses the average intensities in a circular wavefront shaping region with an 8 pixel radius, $I_{\text{wfs}}$, as well as the average intensity around the circle in a 180-by-180 pixel square, $I_{\text{bg}}$. At each step during a measurement or a simulation, these intensities are measured at each phase shift of 160 randomly selected SLM pixels. So, we measure 
$I_{\text{wfs}}(\phi)$ and $I_{\text{bg}}(\phi)$, where $\phi$ is the phase shift set to the 160 random pixels. As mentioned in the main text, only optimizing $I_{\text{wfs}}$ yielded poor contrast. We therefore define the intensity ratio $R(\phi) = I_{\text{wfs}}(\phi)/I_{\text{bg}}(\phi)$. For the first 1000 steps, the algorithm only optimizes $I_{\text{wfs}}$ to increase the signal. Thereafter, at each algorithm step $i$, the optimization parameter is given by the ``weighted'' multiplication
\begin{equation}\label{eq:opt_param}
    O_i(\phi) = \left(I_{\text{wfs}}(\phi)\right)^{f(i')}\left(R(\phi)\right)^{g(i')},
\end{equation}
where $i' = i - 1000$. At each step, a sine is fitted through $O_i(\phi)$ in order to find the optimum phase shift for the 160 selected pixels. At the end of the step, these pixels are then updated with this optimum phase shift and the next step starts. The two exponent functions $f$ and $g$ are decaying and growing functions, respectively, taken as
\begin{equation}\label{eq:opt_expo}
    \begin{aligned}
        f(i') &= 1 - e^{-i'/250}, \\
        g(i') &= e^{-i'/1000}.
    \end{aligned}
\end{equation}
The transition from intensity optimization to ratio optimization is smooth and takes many steps. This is done because the ratio might still be very low after the first 1000 steps, which makes suddenly optimizing the ratio hard. The values of 250 and 1000 were based on experimentation. Faster switching (i.e. lowering the values 250 and 1000) meant that sometimes the optimization failed completely, and slower switching (i.e. increasing the values 250 and 1000) means having to wait longer for the optimum in the ratio. After about 3000 steps ($i' = 2000$), the optimization parameter is mostly based on the ratio. Note that Eqs. (\ref{eq:opt_param}) and (\ref{eq:opt_expo}) were also used in the simulation.

\begin{figure}[htbp]
	\centering
	\includegraphics[width=5in]{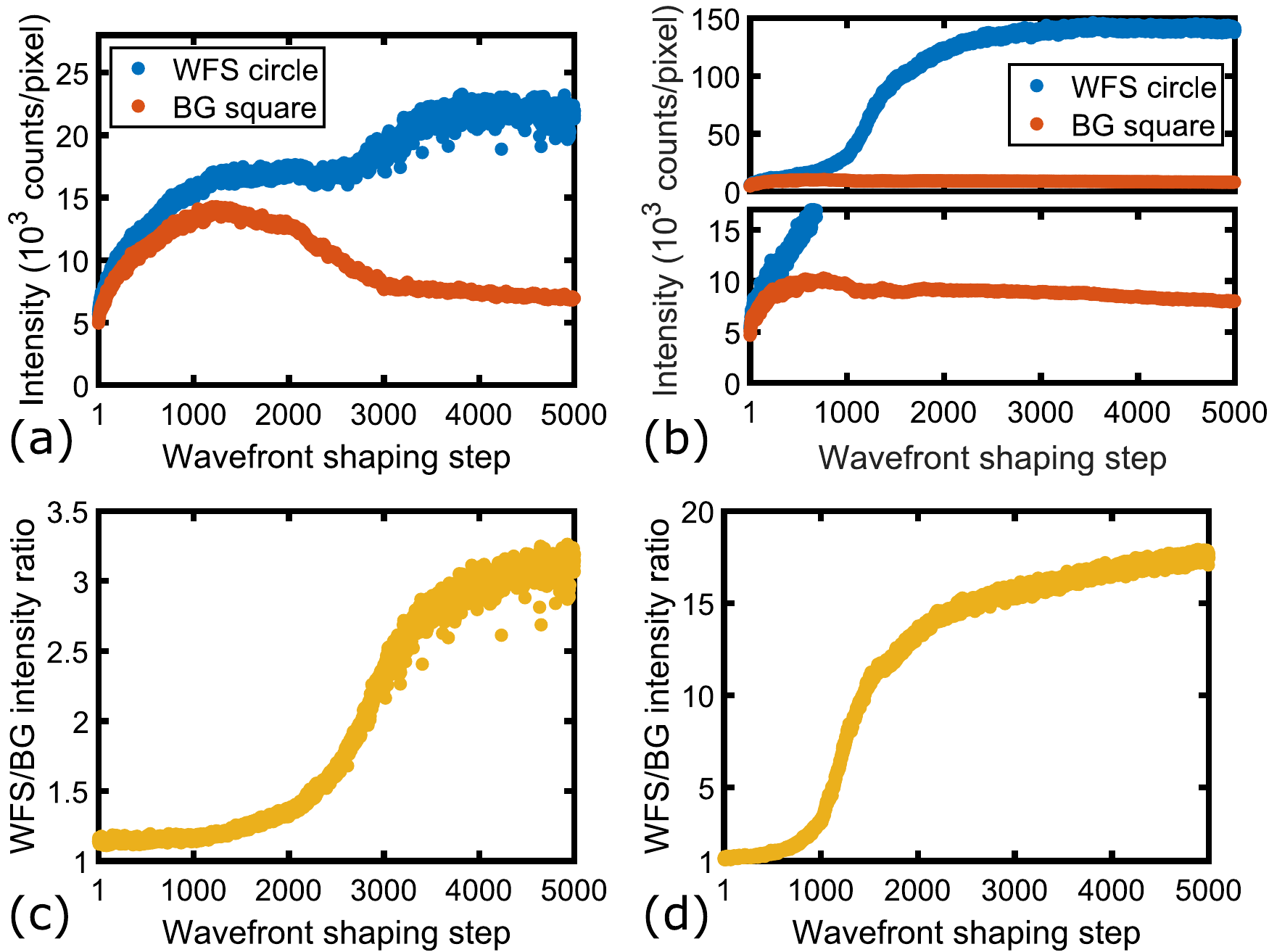}
	\caption{(a) Average intensity in the wavefront shaping (WFS) circle and the background (BG) square in the experiment. (b) Average intensity in the wavefront shaping (WFS) circle and the background (BG) square in the simulation. The bottom panel shows a vertically zoomed section of the upper panel to show the evolution of the background intensity better. (c) The ratio between the WFS and BG intensities for the experimental data in (a). (d) The ratio between the WFS and BG intensities for the simulated data in (b).}
	\label{fig:progress}
\end{figure}

Fig. \ref{fig:progress} shows the progress of a single experimental and simulated wavefront shaping run, taken from the center spots of the experimental and simulated composite grids (see Fig. \ref{fig:composite_grid}). The simulated photon number has been scaled up by a factor of 2 to account for the excess noise due to the EM process and by an additional factor of 500 real gain and is plotted Fig. \ref{fig:progress}(b). Poisson noise was applied before this scaling, so that the Poisson noise in the simulation due to the finite number of photons should be comparable to the experiment. Both the background and wavefront shaping region intensity grow in the first 1000 steps. After the transition to optimization of the ratio occurs, however, the background intensity stops growing and even decreases again. Both the experiment and the simulation show this behaviour. The ratios grow more rapidly at that point, but effectively reach a plateau before the algorithms are terminated after 5000 steps. However, the simulation reaches a much higher ratio, because the increase in the average intensity in the wavefront shaping region is much larger. This is reflected in the fact that the simulation can reach much higher contrast than the experiment.

\section*{Funding}
Funding is acknowledged from the Nederlandse Wetenschaps Organisatie (NWO) via QuantERA QUOMPLEX (Grant No. 680.91.037), and NWA (Grant No. 40017607). 

\section*{Acknowledgments}
We thank Lars van der Hoeven for help with initial measurements, Boris Škorić for discussions and Willem Vos for support.

\section*{Disclosures}
The authors declare no conflicts of interest.

\bibliography{tWFS_VelsinkV2.bib}

\end{document}